\documentclass[transmag]{IEEEtran}
\usepackage{latexsym}
\usepackage{graphicx}
\usepackage{amsfonts,amssymb,amsmath}
\usepackage{hyperref}
\usepackage{dsfont}

\def\BibTeX{{\rm B\kern-.05em{\sc i\kern-.025em b}\kern-.08em T\kern-.1667em\lower.7ex\hbox{E}\kern-.125emX}}
\begin{document}

\title{State-Space Models Win the IEEE DataPort Competition on Post-covid Day-ahead Electricity Load Forecasting}

\author{Joseph de Vilmarest and Yannig Goude
\thanks{J. de Vilmarest is with the Laboratoire de Probabilit\'es, Statistique et Mod\'elisation, Sorbonne Universit\'e and with \'Electricit\'e de France R\&D (e-mail: joseph.de\_vilmarest@sorbonne-universite.fr).}
\thanks{Y. Goude is  with \'Electricit\'e de France R\&D and the Laboratoire de Math\'ematique d'Orsay, Universit\'e Paris-Saclay (e-mail: yannig.goude@edf.fr).}}

\IEEEtitleabstractindextext{
\begin{abstract}
We present the winning strategy of an electricity demand forecasting competition. This competition was organized to design new forecasting methods for unstable periods such as the one starting in Spring 2020. We rely on state-space models to adapt standard statistical and machine learning models. We claim that it achieves the right compromise between two extremes. On the one hand, purely time-series models such as autoregressives are adaptive in essence but fail to capture dependence to exogenous variables. On the other hand, machine learning methods allow to learn complex dependence to explanatory variables on a historical data set but fail to forecast non-stationary data accurately. The evaluation period of the competition was the occasion of trial and error and we put the focus on the final forecasting procedure. In particular, it was at the same time that a recent algorithm was designed to adapt the variances of a state-space model and we present the results of the final version only. We discuss day-to-day predictions nonetheless.
\end{abstract}

}

\maketitle

\section{Introduction}
\IEEEPARstart{E}{lectricity} demand forecasting is a crucial task for grid operators. Indeed the production must balance the consumption as storage capacities are still negligible compared to the load. Time series methods have been applied to address that problem, relying on calendar information and lags of the electricity consumption. Statistical and machine learning models have been designed to use exogenous information such as meteorological forecasts (the load usually depends on the temperature for instance, due to electric heating and cooling).

The field has been thoroughly studied the past decades and we will not propose here an exhaustive bibliographic study. Instead, we choose  to focus on recent results in the different forecasting challenges related  to this the field. The Global Energy Forecasting Competitions (GEFCOM) (\cite{hong2014global}, \cite{HONG2016896} and \cite{hong2019global}) provide a large benchmark of popular and efficient load forecasting methods. Black box machine learning models such as gradient boosting machines \cite{lloyd2014gefcom2012} and neural networks \cite{ryu2017deep, dimoulkas2019neural} rank among the first as  well as statistical models like Generalized Additive Models (GAM) \cite{nedellec2014gefcom2012, dordonnat2016gefcom2014} or parametric regression models \cite{CHARLTON2014364, ZIEL20191400}. Ensemble methods or expert aggregation are also a common practice for competitors \cite{gaillard2016additive, smyl2019machine}.

The behaviour of the consumption changed abruptly during the coronavirus crisis, especially during lockdowns imposed by many governments. These changes of consumption mode have been challenging for electricity grid operators as historical forecasting procedures performed poorly. It is to be noted that purely time series methods like autoregressives didn't drift as they are very adaptive in essence, but they fail to capture the dependence of the load to, for instance, meteorological variables. Therefore designing new forecasting strategies to take that evolution into account is important to reduce the cost of forecasting errors and to ensure the stability of the network in the future.

We claim that state-space models allow the best of both worlds. First, machine learning models trained on historical data are used to design new feature representations. Second, a state-space representation yields a methodology to adapt these complex forecasting models.

Our work extends a previous study on the French electricity load \cite{obst2021adaptive} where a state-space approach to adapt generalized additive models was presented.
The novelty of this article lies both in the method and in the application. First, besides generalized additive models we apply our procedure to other widely used machine learning models including neural networks. Second, after applying a standard Kalman filter we apply VIKING (Variational Bayesian Variance Tracking), another state-space approach allowing to estimate jointly the state and the variances \cite{tsw}. Third, our procedure resulted in the winning strategy in a competition on post-covid day-ahead electricity demand forecasting\footnote{\href{https://ieee-dataport.org/competitions/day-ahead-electricity-demand-forecasting-post-covid-paradigm}{https://ieee-dataport.org/competitions/day-ahead-electricity-demand-forecasting-post-covid-paradigm}}, motivating the efficiency of the proposed approach.

Section \ref{sec:data_pres} is an introduction to the competition and we detail how we handled the data provided. In Section \ref{sec:timeinvariant} we discuss the meteorological variables and we present standard forecasting methods. The core of our strategy is Section \ref{sec:adaptation} where we propose a generic state-space framework to adapt these methods. We discuss the numerical performances of the various models in Section \ref{sec:experiments} and we combine them through aggregation of experts to leverage each model's advantages.

\section{Data Presentation and Pre-Processing}\label{sec:data_pres}
\begin{figure}
    \centering
    \includegraphics[width=8cm]{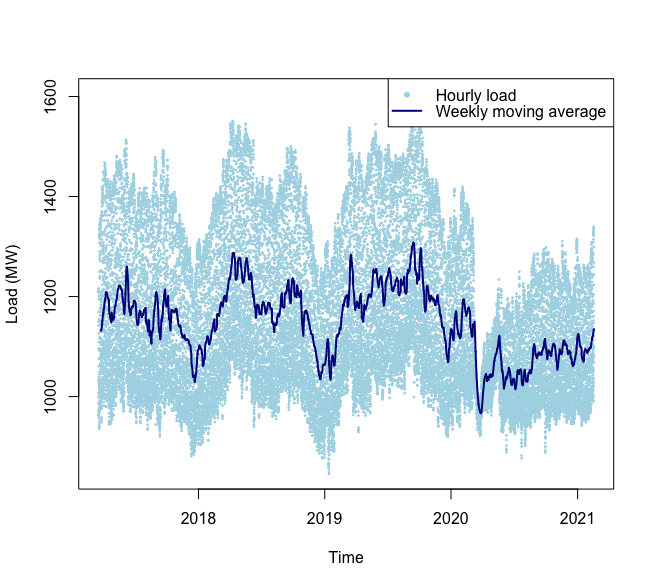}
    \includegraphics[width=8cm]{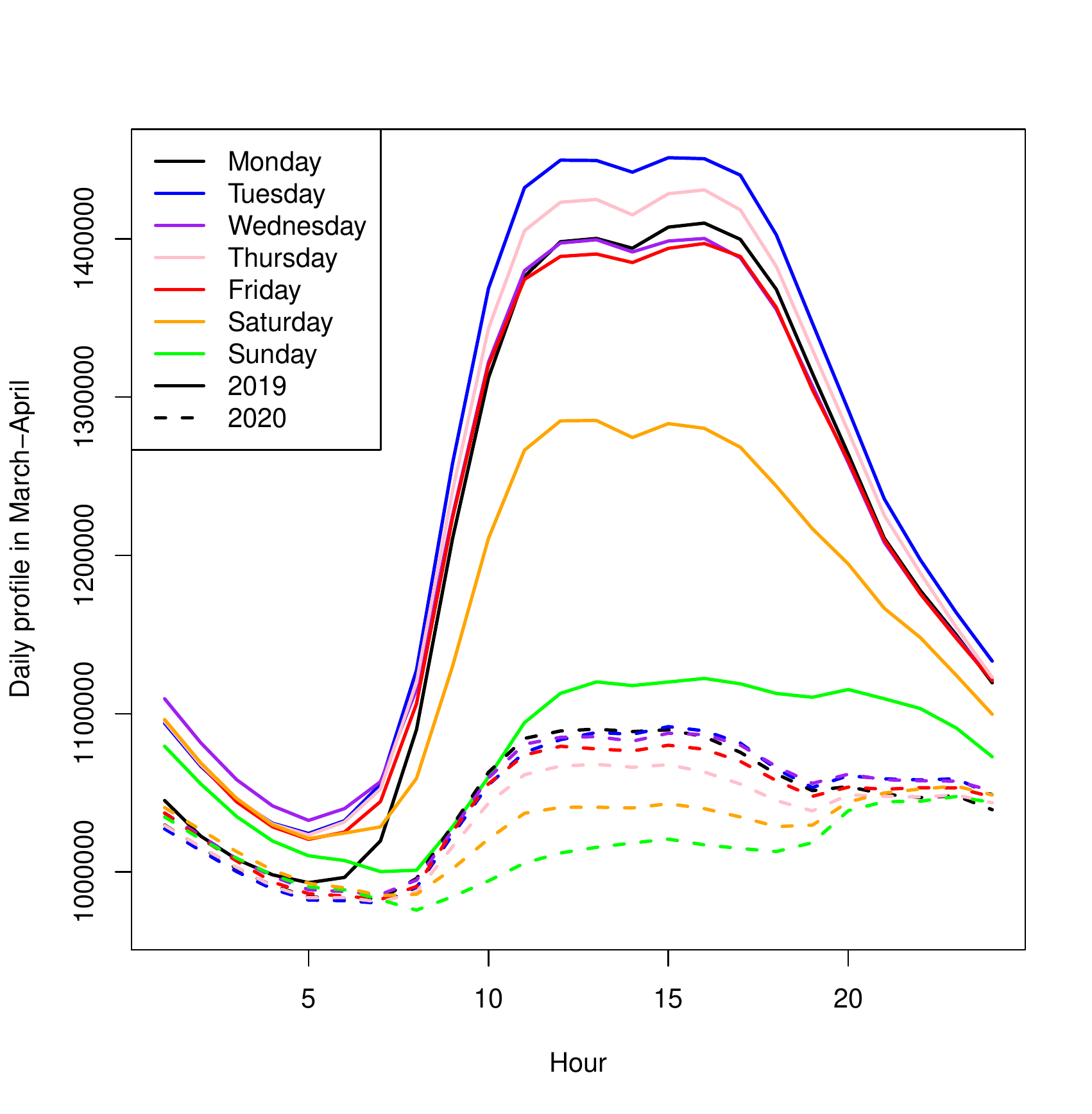}
    \caption{On top: electricity load from March 18\textsuperscript{th} 2017 to February 16\textsuperscript{th} 2021. On the bottom: daily profiles of the electricity load in March-April 2019 compared to March-April 2020.}
    \label{fig:daily_profile}
\end{figure}

The objective of the competition was to predict the electricity load of an undisclosed location of average consumption 1.1 GW, that is of the order of one million people in western countries. The break in the electricity demand in March 2020 is clear in Figure \ref{fig:daily_profile}. The objective of the competition was to design new strategies for day-ahead forecasting in order to be robust to this unstable period.

\subsection{Time segmentation}
The competition's setting was to forecast the hourly load 16 to 40 hours ahead in an online manner. Precisely we had to predict the consumption of each hour of day d with data up to 8AM day d-1. After our prediction was sent a new batch of data up to 8AM day d was released so that we had to predict day d+1 ...

The evaluation was based on the Mean Average Error on the period ranging from January 18\textsuperscript{th} to February 16\textsuperscript{th} 2021. To build a forecasting model, the historical load starting from March 18\textsuperscript{th} 2017 was provided, as well as meteorological forecasts and realizations during the same period.

\subsection{Meteorological Forecasts}
Aside from calendar variables, is is usual that the most important exogenous factor explaining the electricity demand is meteorology. The dependence of the load to the temperature for example is due to electric heating and cooling. Moreover, the dependence of the electricity demand to meteorology is augmented by the development of decentralized renewables. Indeed, small renewable production is often used by its owner, yielding a net consumption that highly depends on wind or solar radiation.
Therefore the error of a forecasting model for the electricity demand crucially depends on the performance of the meteorological forecasts.

The data of the competition includes forecasts and realization of the temperature, the cloud cover, the pressure, the wind direction and speed. These forecasts are assumed to be known 48 hours in advance and invariant after, thus they can be used to forecast the load at the 16 to 40 hours horizon.

\begin{figure}
    \centering
    \includegraphics[width=4.3cm]{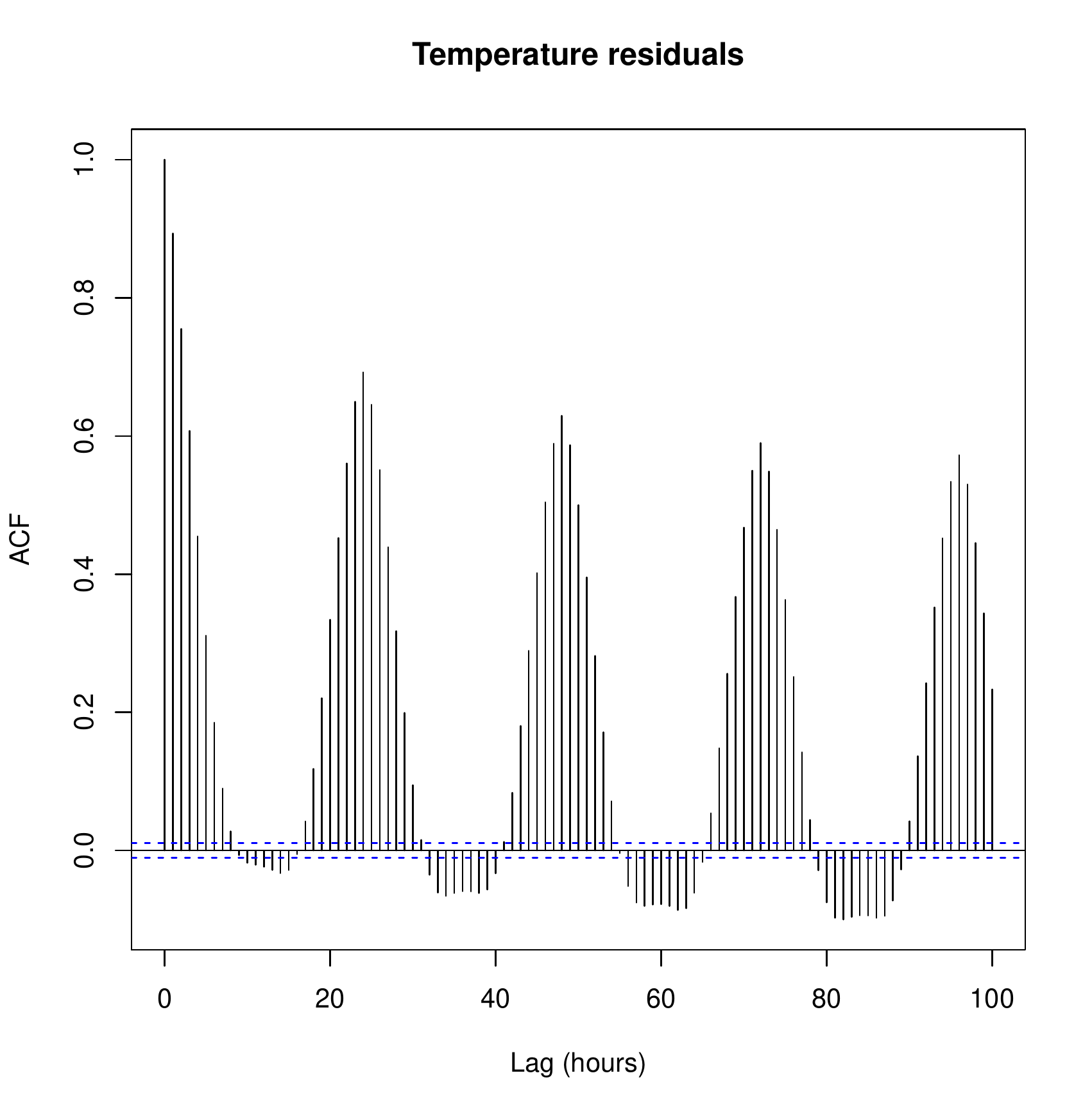}
    \includegraphics[width=4.3cm]{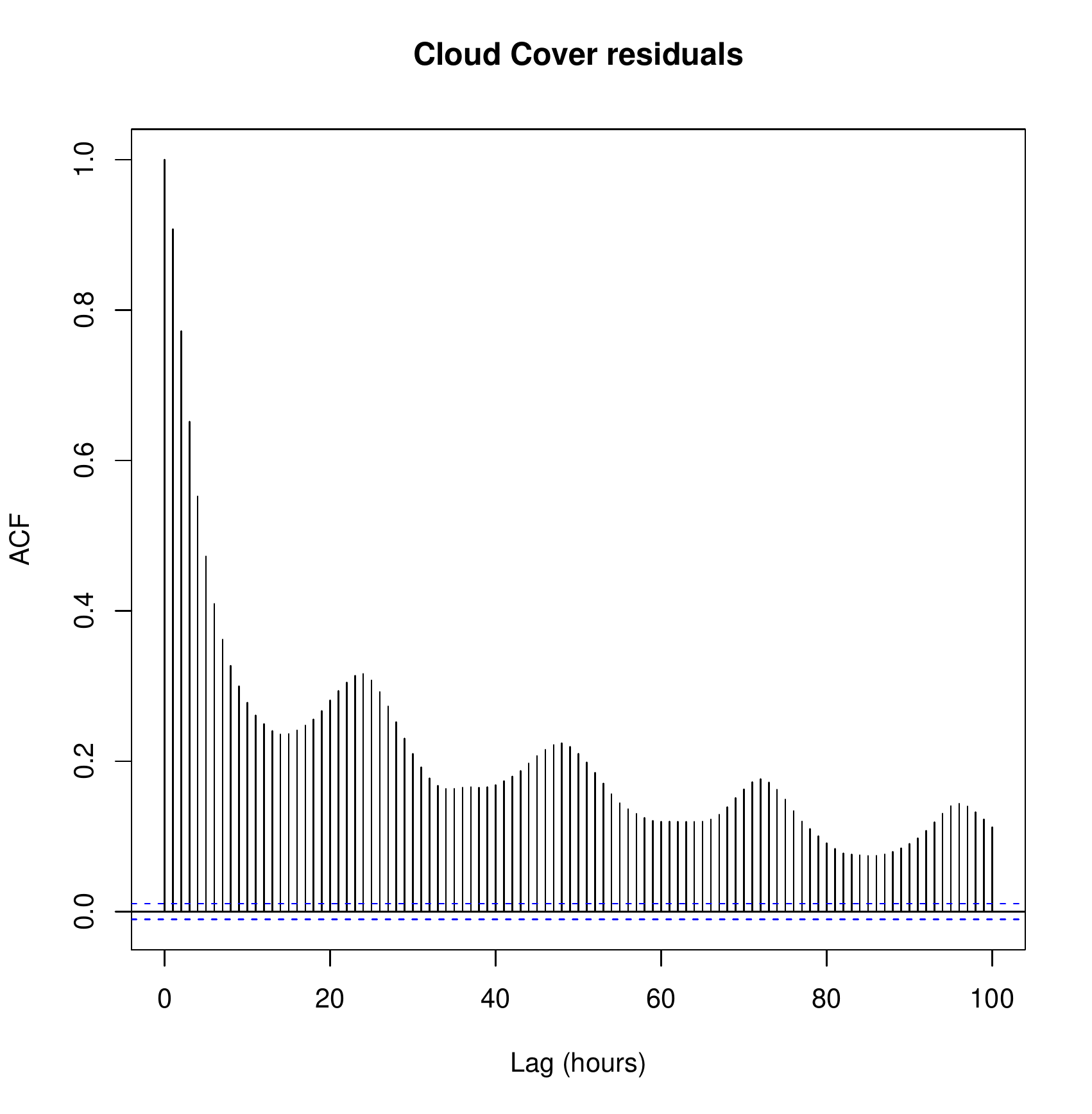}
    \includegraphics[width=4.3cm]{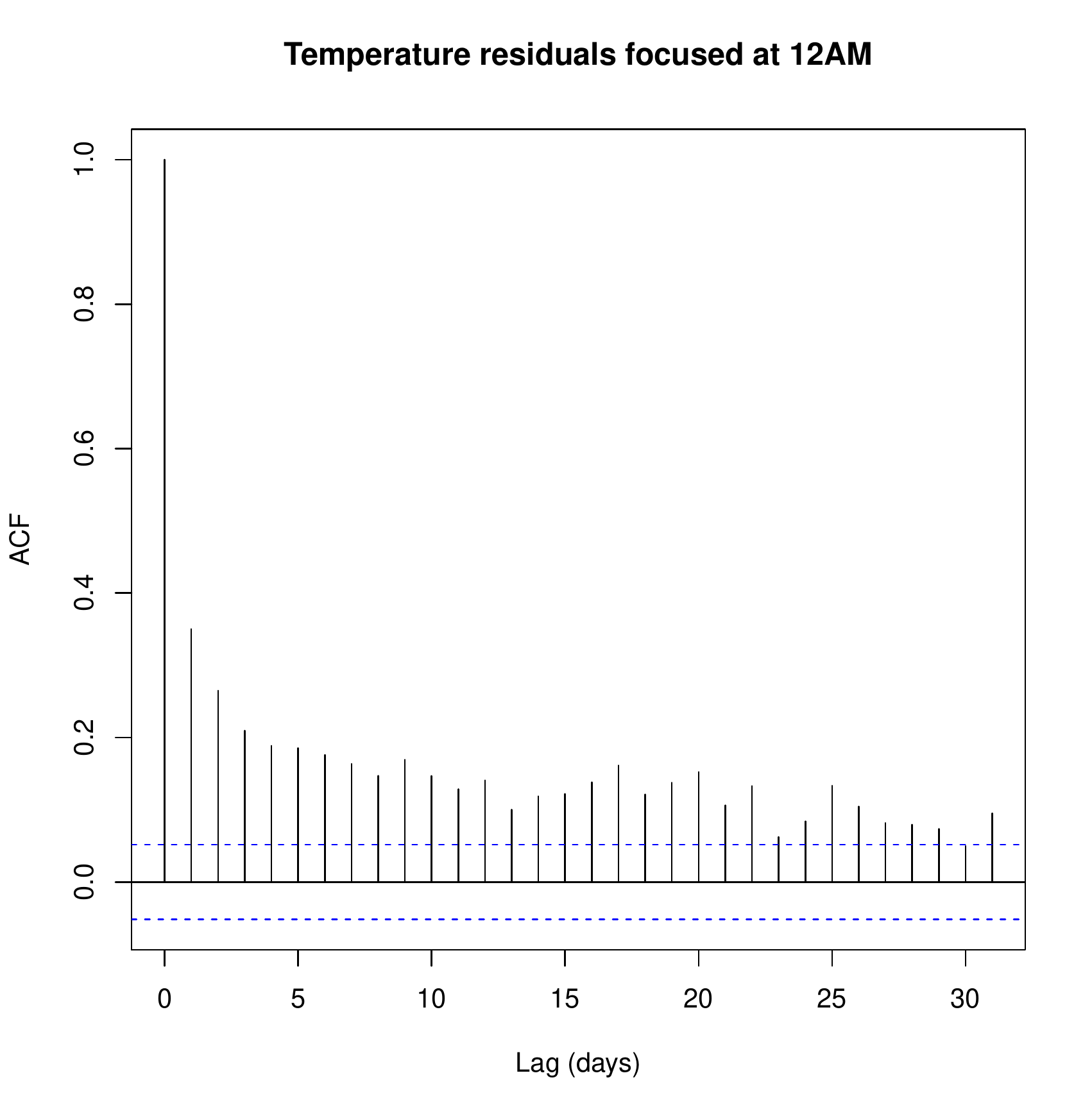}
    \includegraphics[width=4.3cm]{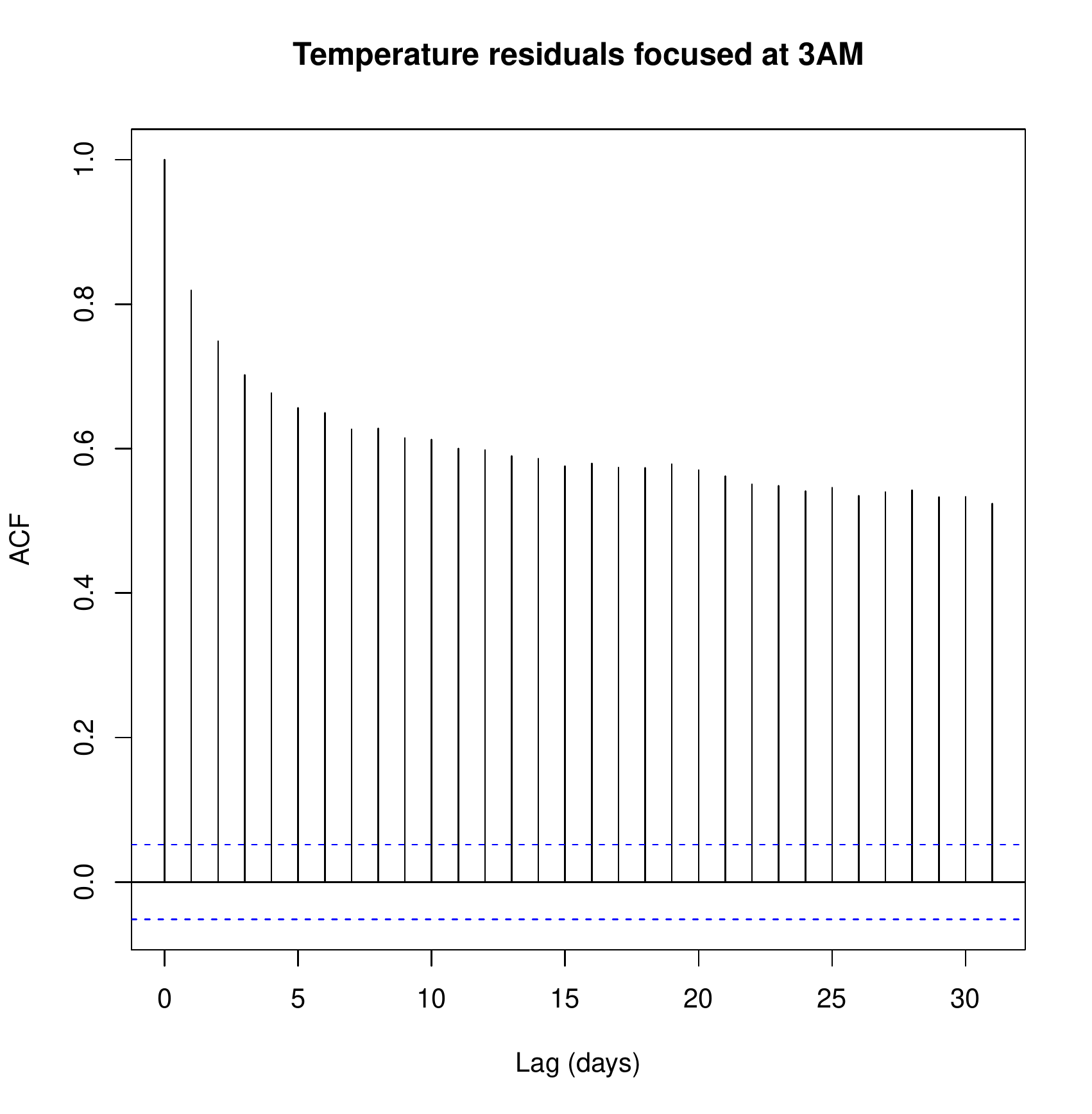}
    \caption{On top: auto-correlation plots of the temperature (left) and cloud cover (right) forecasting residuals, with lags in hours. On the bottom: auto-correlation plots of the temperature residuals focused on a specific hour of the day (midnight on the left, 3AM on the right), with lags in days.}
    \label{fig:meteoresiduals}
\end{figure}

However from the statistical properties of the meteorological forecasting residuals (c.f. Figure \ref{fig:meteoresiduals}), we conjecture that the forecasts come from physical models that need to be statistically corrected. Indeed, as the forecasts are available 48 hours in advance, if a statistical correction had been applied then auto-correlations of the residuals over 48 hours would be negligible.
We thus use correction models close to autoregressives on the residuals.

Formally, let $(z_t)$ be any of the meteorological variable and $(\hat z_t)$ the forecast given in the data set. Then we use the model
\begin{align*}
    z_t = \alpha \hat z_t + \sum\limits_{l\in \mathcal{L}_{p,P,h(t)}} \beta_l (z_{t-l} - \hat z_{t-l}) + \gamma z_{t-l_0(t)} +\delta + \varepsilon_t \,,
\end{align*}
where $h(t)\in\{0,\hdots,23\}$ is the hour of the day of time $t$ and
\begin{align*}
    & l_0(t) = 
    \begin{cases}
    24 & \text{if } h\le 7\,, \\
    48 & \text{if } h>7\,,
    \end{cases} \\
    & \mathcal{L}_{p,P,h} = 
    \begin{cases}
    \resizebox{0.52\hsize}{!}{$\{24,\hdots,24*P,h+17,\hdots,h+16+p\}$} & \text{if } h\le 7\,, \\
    \resizebox{0.6\hsize}{!}{$\{48,\hdots,24*(P+1),h+17,\hdots,h+16+p\}$} & \text{if } h>7\,.
    \end{cases}
\end{align*}
In other words, we forecast the residual of the variable of interest with a linear model on
\begin{itemize}
    \item 
    the last $P$ available daily lags of the residual,
    \item
    the last $p$ available lags of the residual (up to 7AM of the previous day),
    \item
    the forecast,
    \item
    the last daily lag of the variable of interest.
\end{itemize}

We optimize the coefficients separately for each hour of the day for the temperature, whereas we use the same coefficients at each hour of the day for the cloud cover, pressure and wind speed (except the intercept term). We don't correct the wind direction. The parameters $p$ and $P$ are selected based on BIC. We display in Table \ref{tab:weather_correction} the error of the initial forecast, compared to simply using the last daily lag of the variable of interest, and our corrected forecast.

\begin{table}
    \begin{center}
    \caption{Mean Average Error of different meteorological forecasts.}
    \label{tab:weather_correction}
    \begin{tabular}{|c c c c|}
        \hline
        & Initial & Last daily lag & Corrected \\
        \hline
        Temperature ($^\circ$C) & 3.00 & 2.11 & 1.69 \\
        \hline
        Cloud cover (\%) & 17.28 & 18.74 & 14.99 \\
        \hline
        Pressure (kPa) & 0.506 & 1.30 & 0.423 \\
        \hline
        Wind Speed (km/h) & 4.53 & 3.49 & 2.53 \\
        \hline
    \end{tabular}
    \end{center}
    The first column is the forecast given in the data set. The second one consists in using the variable of interest with a 24 or 48-hour delay. The last is our corrected forecast. We evaluate through the mean average error during 2020 while we train the corrections on the data before 2020.
\end{table}

\section{Time-invariant Experts}\label{sec:timeinvariant}
We summarize the explanatory variables used in our forecasting models:
\begin{itemize}
    \item 
    calendar variables: the hour of the day, the day of the week, the time of year ($Toy$) growing linearly from 0 on January 1\textsuperscript{st} to 1 on December 31\textsuperscript{st}, and a variable growing linearly with time to account for a trend,
    \item
    meteorological forecasts after statistical correction: the temperature along with exponential smoothing variants of parameters 0.95 and 0.99 (respectively $Temps95$ and $Temps99$), the cloud cover, the pressure, the wind direction and speed,
    \item
    lags of the electricity load: the load a week ago $LoadW$ and the last load available $LoadD$ (a day ago for the forecast before 8AM and two days ago after 8AM, this constraint coming from the availability of the online data during the competition).
\end{itemize}

\begin{figure}
    \centering
    \includegraphics[width=4.3cm]{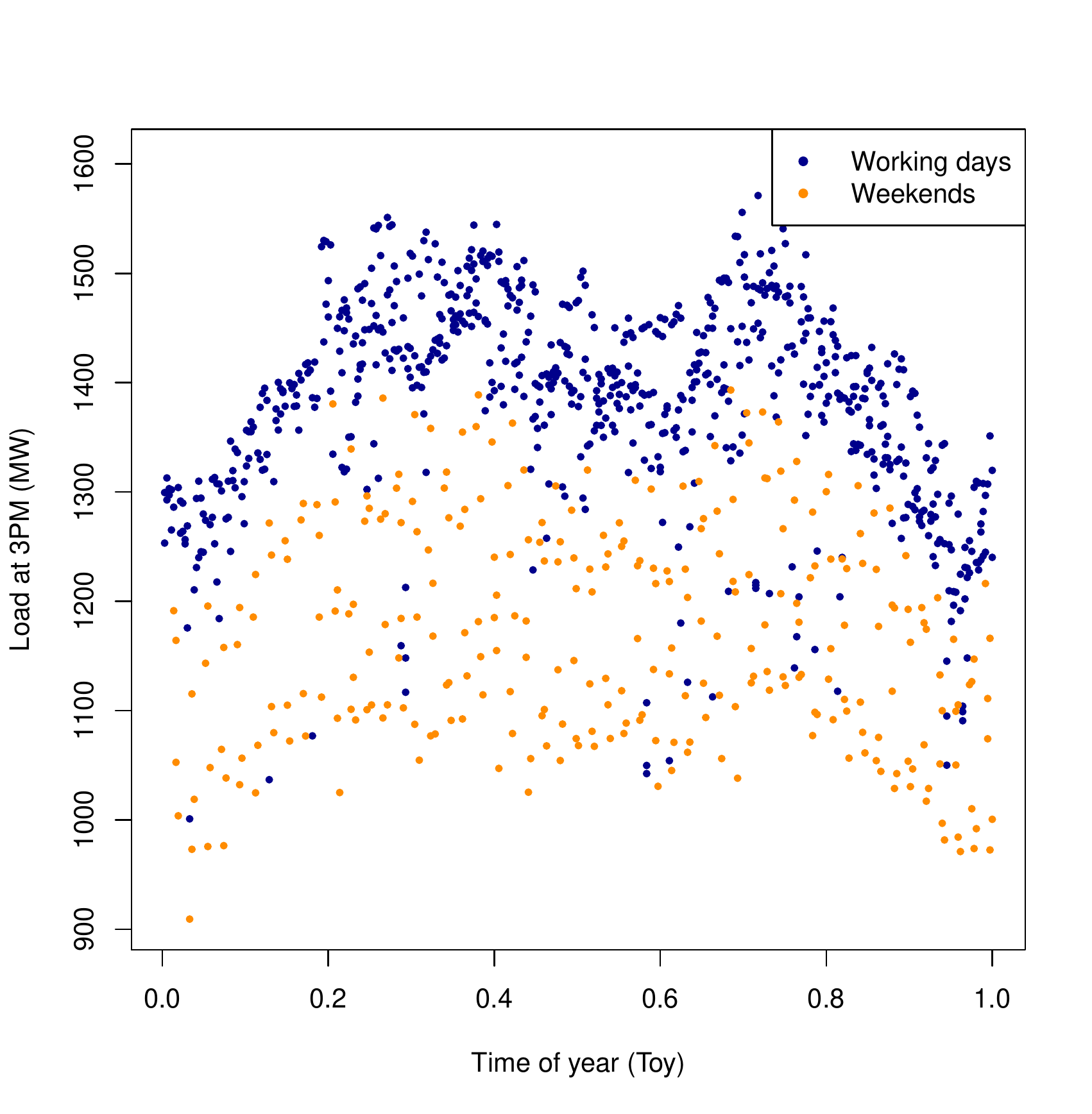}
    \includegraphics[width=4.3cm]{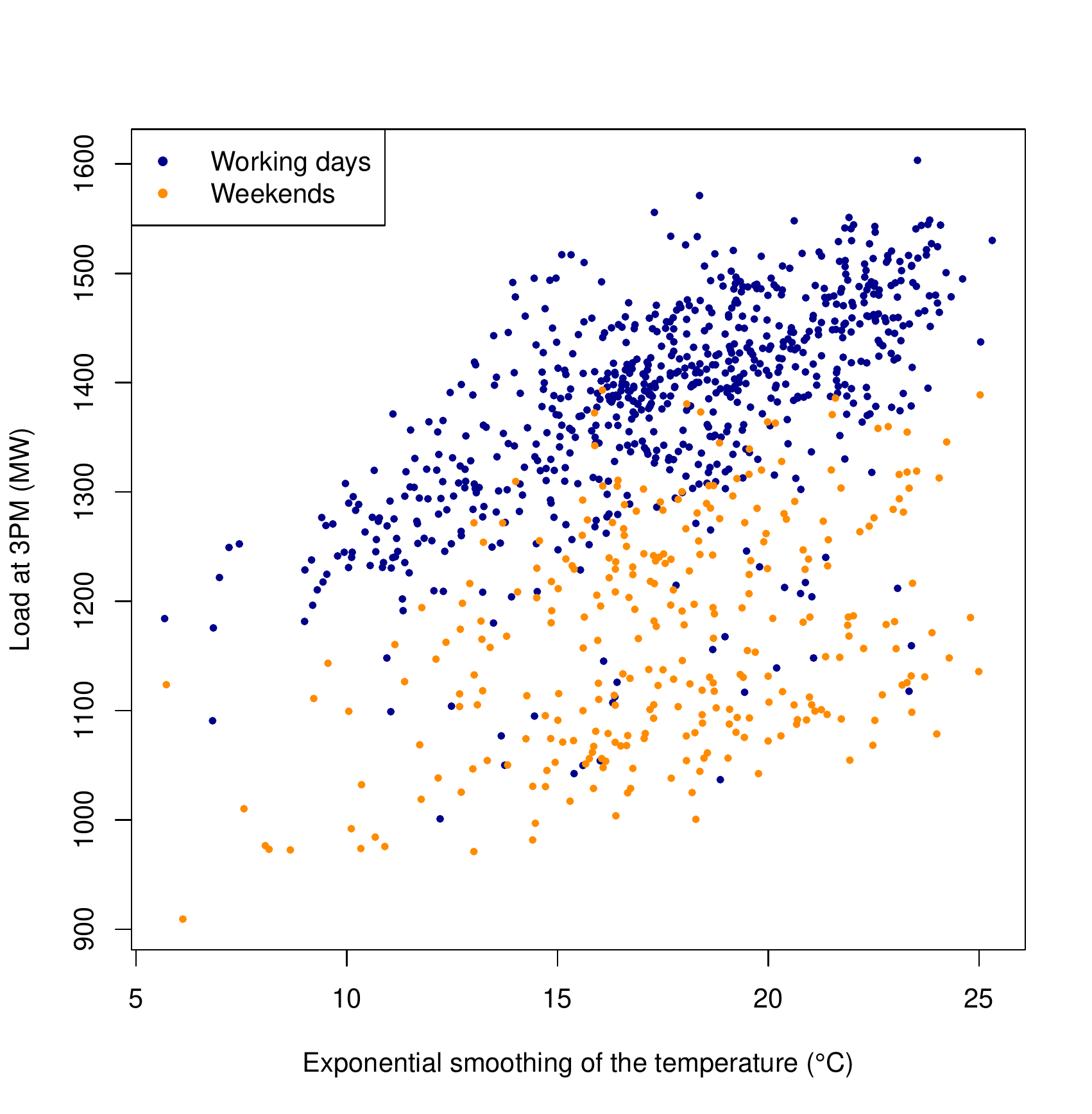}
    \includegraphics[width=4.3cm]{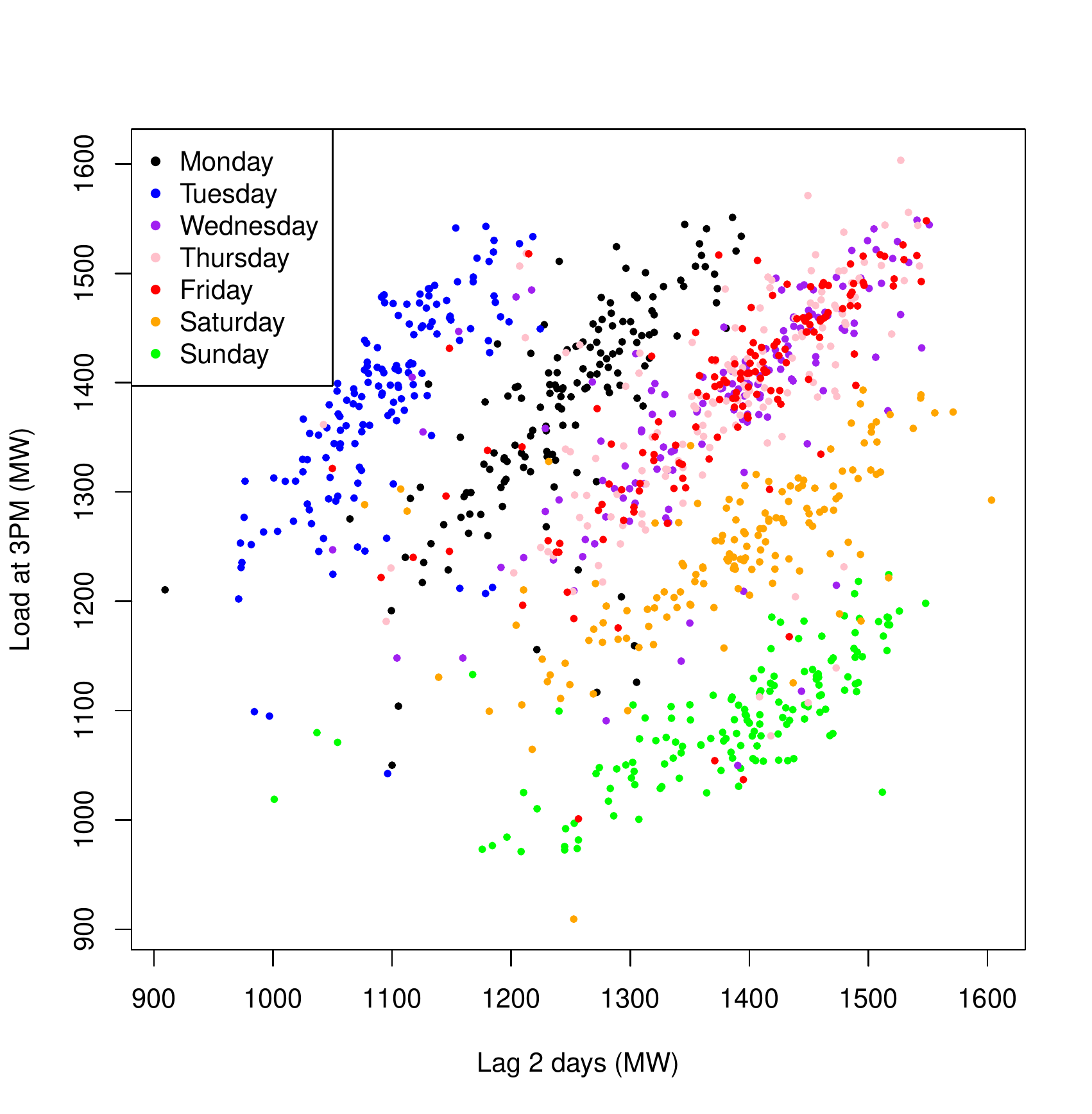}
    \includegraphics[width=4.3cm]{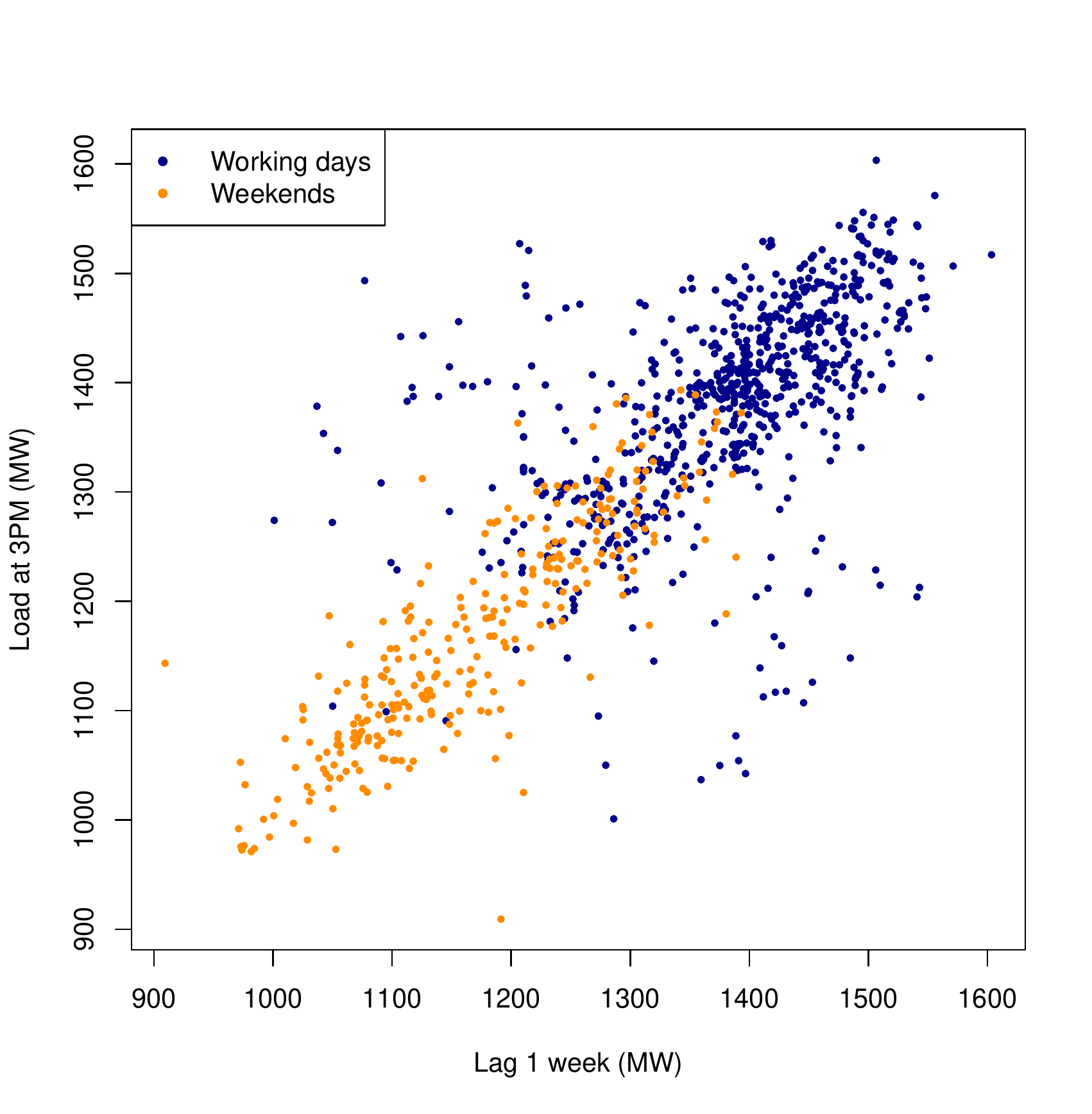}
    \caption{Dependence of the load at 3PM to different covariates on the data up to January 1\textsuperscript{st} 2020.}
    \label{fig:explore}
\end{figure}

The dependence to the hour of the day and the day of the week is well observed in Figure \ref{fig:daily_profile}. We display in Figure \ref{fig:explore} the dependence of the load to a few of the aforementioned covariates.

\subsection{Statistical and Machine Learning Methods}\label{sec:statsml}
Based on the covariates we experiment a few classical predictive models. We define independent models for the different hours of the day as is usual in electricity load forecasting. For each model we use the same structure for the different hours but we learn the model parameters independently for each time of day based on the training data of that particular time of day. In what follows we denote by $y_t$ the load at time $t$.

\begin{itemize}
\item
{\bf Autoregressive}. We consider a seasonal autoregressive model based on the daily and weekly lags of the load:
\begin{align}
    & \label{eq:autoregressive}
    y_t = \sum\limits_{l\in\mathcal{L}_{h(t)}} \alpha_{l} y_{t-l} + \sum\limits_{1\le l\le 6} \alpha_{7\times24l} y_{t-7\times24l} + \varepsilon_t\,,\\
    \nonumber & \mathcal{L}_{h} = \begin{cases}
    \{24,48,72\}\qquad \text{if } h\le 7 \,,\\
    \{48,72,96\}\qquad \text{if } h> 7\,.
    \end{cases}
\end{align}

\item
{\bf Linear regression}. We use a linear model with the following variables: temperature, cloud cover, pressure, wind direction and speed, day type (7 booleans), time of year, linear trend variable, and the two lags $LoadW$ and $LoadD$.

\item
\textbf{Generalized additive model (GAM)}. We propose a Gaussian generalized additive model \cite{wood2017generalized}:
\begin{align*}
    y_t & = \sum_{i=1}^6 \beta_i \mathds{1}_{DayType_t = i} + \gamma Temps95_t + f_1(Toy_t) \\
    & \quad + f_2(LoadD_t) + f_3(LoadW_t) + \alpha t + \beta_0 + \varepsilon_t \,,
\end{align*}
where $f_1$ is obtained by penalized regression on cubic cyclic splines and $f_2,f_3$ on cubic regression splines.

\item
\textbf{Random Forest (RF)}. We build a random forest \cite{RF} with the following covariates: linear trend variable, time of year, day type, the two lags and the two exponential smoothing variables of the temperature. Quantile variants were also computed.
% (denoted by {\bf RF40, RF50, RF60}).

\item
\textbf{Random Forest (RF\_GAM)}. We also correct the GAM using a random forest on the GAM residuals, with the same covariates as in {\bf RF} to which we add the GAM effects $f_1(Toy_t)$, $f_2(LoadD_t)$, $f_3(LoadW_t)$ as well as lags (one week, one or two days) of the GAM residuals.

\item
\textbf{Multi-Layer Perceptron (MLP)}. Finally we test a multi-layer perceptron of 2 hidden layers of 15 and 10 neurons using hyperbolic tangent activation. We take as input: linear trend variable, time of year, day type, the exponential smoothing variable $Temps95$ and the two lags.
\end{itemize}

\subsection{Intraday correction}\label{sec:intraday}
Although it performs better to use different models at the different hours of the day, let it be noted that the correlation between different hours is important. To capture intraday information we fit on the residuals of each model an autoregressive model incorporating lags of the 24 last available hours and optimized for each forecast horizon. This follows from the intuition that to predict the load at 8AM, instead of using as the last available data a delay of 48 hours, we can use a 25-hour delay.

We apply this correction to the models presented in Section \ref{sec:statsml} as well as to the ones resulting from the adaptation framework of Section \ref{sec:adaptation}, see the numerical improvements in Section \ref{sec:exp_indiv}.

\section{Adaptation using State-Space Models}\label{sec:adaptation}
\begin{figure}
    \centering
    \includegraphics[width=8cm]{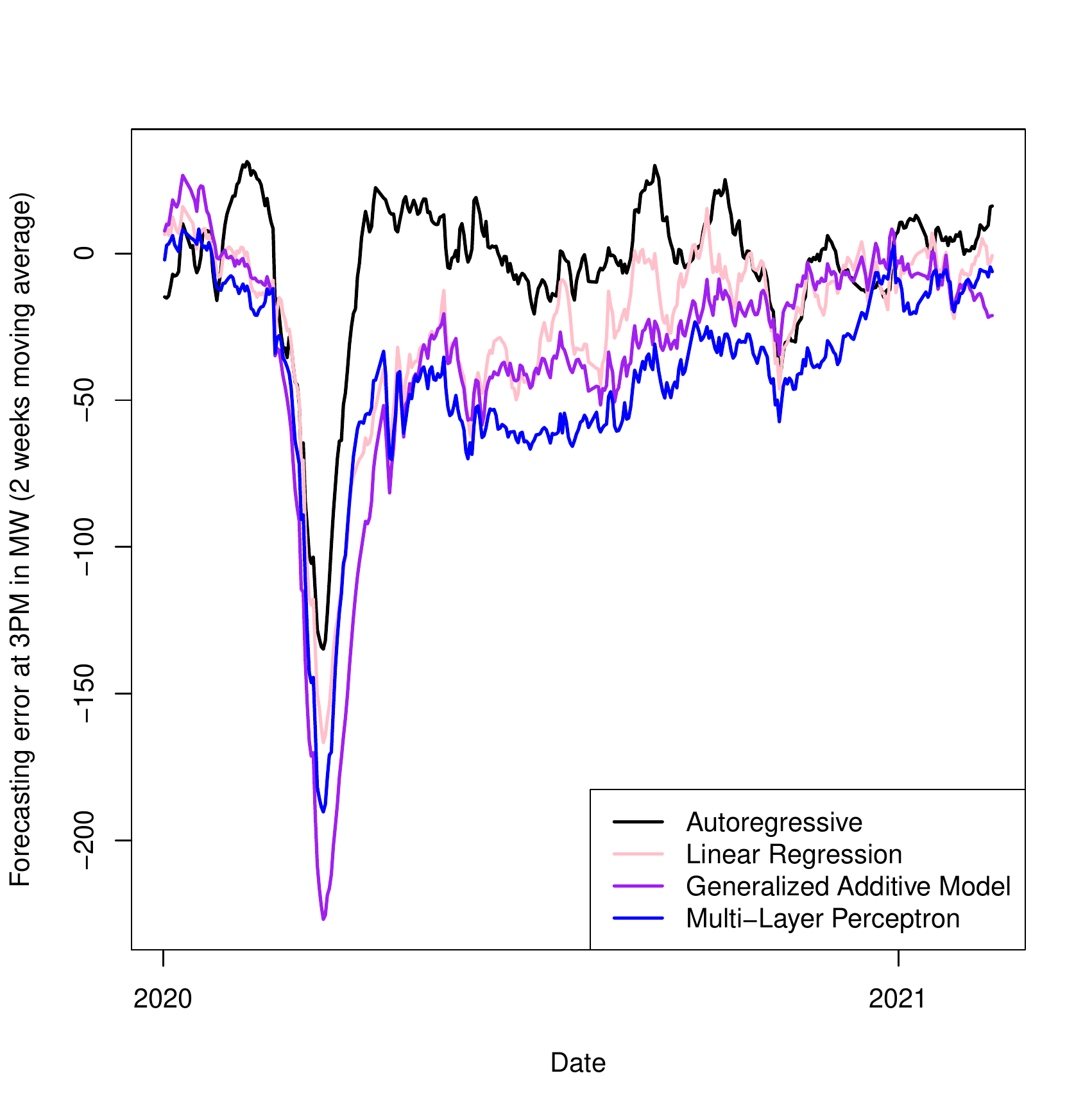}
    \caption{Evolution of the forecasting error for the different models introduced in Section \ref{sec:statsml} trained on the data up to January 1\textsuperscript{st} 2020 and with intraday correction (Section \ref{sec:intraday}).}
    \label{fig:adaptation_motivation}
\end{figure}

Due to the lockdowns the consumer's behaviours changed abruptly and therefore the models presented in Section \ref{sec:statsml} perform poorly during Spring 2020 and afterwards, see Figure \ref{fig:adaptation_motivation}.
To adapt the models in time, we rely on linear gaussian state-space models, summarized as
\begin{align*}
    & \theta_t - \theta_{t-1} \sim \mathcal{N}(0,Q_t)\,, \\
	& y_t - \theta_t^\top x_t \sim \mathcal{N}(0,\sigma_t^2) \,,
\end{align*}
where $\theta_t$ is the latent state, $Q_t$ the process noise covariance matrix and $\sigma_t^2$ is the observation variance.

\subsection{Definition of $x_t$}
\begin{figure}
    \centering
    \includegraphics[width=6cm]{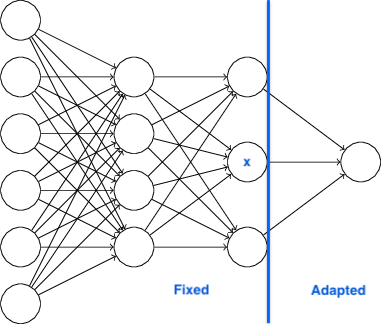}
    \caption{Diagram of the definition of the features to adapt the MLP. The network has two hidden layers of 15 and 10 neurons, we freeze all the weights except the last ones.}
    \label{fig:mlp}
\end{figure}

This state-space representation is natural for linear regression for which $x_t$ is the vector containing the explanatory variables detailed in Section \ref{sec:statsml}. Autoregressive models also fit directly in that framework, as they are in fact linear models on lags of the load, see Equation \eqref{eq:autoregressive}. To adapt GAM and MLP we linearize the models and $x_t$ is just another feature representation. We freeze the non-linear effects in the GAM as in \cite{obst2021adaptive}, and $x_t$ contains the different effects, linear and non-linear. We apply a similar approach for the MLP, for which we freeze the deepest layers and we learn the last one, that is $x_t$ is the final hidden state, see Figure \ref{fig:mlp}.

The state-space approach is not applied to the random forest. For the latter we compare with incremental offline random forests, consisting in re-training the random forest each day with all the data available at the time.

\subsection{Kalman Filter}\label{sec:kf}
Bayesian estimation of the state $\theta_t$ in linear gaussian state-space models is well understood under known variances $\sigma^2,Q$. The best estimator is obtained by the well known Kalman Filter \cite{kalman1961new}.
It yields recursive exact estimation of the mean and covariance matrix of the state given the past observations, denoted respectively by $\hat\theta_t$ and $P_t$. However there is no consensus in the literature as to how to tune the hyper-parameters, see for instance \cite{brockwell1991time,durbin2012time,fahrmeir1992posterior}. The widely used Expectation-Maximization algorithm is an iterative algorithm that guarantees convergence to a local maximum of the likelihood. However there is no global guarantee and in our case it performs poorly. We propose instead the following settings, building on \cite{obst2021adaptive}:

\begin{itemize}
\item
\textbf{Static}. We consider the degenerate setting where $Q_t=0$ and $\hat\theta_1=0,P_1=I,\sigma_t^2=1$.

\item
\textbf{Static break}. We consider a break at March 1\textsuperscript{st} 2020 by setting $\hat\theta_1=0,P_1=I,\sigma_t^2=1,Q_t=0$ except $Q_T=I$ where $T$ is March 1\textsuperscript{st} 2020.

\item
\textbf{Dynamic}. We approximate the maximum-likelihood for constant $\sigma_t^2,Q_t$. We set $P_1=\sigma^2 I$ and we observe that for a given $Q/\sigma^2$ we have closed-form solutions for $\hat\theta_1, \sigma^2$. Then we restrict ourselves to diagonal matrices $Q/\sigma^2$ whose nonzero coefficients are in $\{2^j,-30\le j\le 0\}$ and we apply a greedy procedure: starting from $Q/\sigma^2=0$ we change at each step the coefficient improving the most the likelihood. That procedure is designed to optimize $Q$ on the training data (up to January 1\textsuperscript{st} 2020).

\item
\textbf{Dynamic break}. We use similar $\hat\theta_1,P_1,\sigma_t^2=\sigma^2,Q_t=Q$ as in the dynamic setting except $Q_T=P_1=\sigma^2 I$ where $T$ is March 1\textsuperscript{st} 2020.

\item
\textbf{Dynamic big}. We simply use $\sigma^2=1$ and a matrix $Q$ proportional to $I$ defined based on the 2020 data.
\end{itemize}

Also, it is important that we estimate a gaussian posterior distribution, therefore we have a probabilistic forecast for the load. Precisely, our estimate is $\theta_t\sim\mathcal{N}(\hat\theta_t,P_t)$, thus we have $y_t\sim\mathcal{N}(\hat\theta_t^\top x_t, \sigma^2+x_t^\top P_t x_t)$. The likelihood that is optimized to obtain the dynamic setting is built on that probabilistic forecast of $y_t$ given the past observations.
In the competition we added quantiles of these gaussian distributions as forecasters in an expert aggregation.

\subsection{Dynamical Variances}
The idea behind the break settings introduced in the previous paragraph is that we would like the model to adapt faster during an evolving period such as a lockdown than before. However it consists in modelling a break in the data, a sudden change of state resulting from a noise of much bigger variance at a specific time specified {\it a priori}. A way to extend the approach would be to define a time-varying covariance matrix depending for instance of a lockdown stringency index such as defined by \cite{hale2021global}. However the competition policy forbade the use of external data and the location was undisclosed.

In a more long-term perspective let it be hoped that the evolution of the electricity load won't be driven by lockdowns. It is therefore more generic to learn the variances of the state-space model in an adaptive fashion. We therefore apply a novel approach for time-series forecasting introduced in \cite{tsw}, and named Variational Bayesian Variance Tracking, alias VIKING. We briefly recall how the method works.
This method was designed in parallel of the competition and was improved afterwards. We present the last version only.

We treat the variances as latent variables and we augment the state-space model:
\begin{align*}
	%& \theta_0\sim\mathcal{N}(\hat\theta_0,P_0)\,, \quad 
	%a_0\sim\mathcal{N}(\hat a_0, s_0)\,,\quad 
	%b_0\sim\mathcal{N}(\hat b_0, \Sigma_0)\,,\\
	& a_t - a_{t-1} \sim\mathcal{N}(0,\rho_a)\,, \quad
	b_t - b_{t-1} \sim\mathcal{N}(0,\rho_b)\,, \\
	& \theta_t - \theta_{t-1} \sim\mathcal{N}(0, \exp(b_t) I) \,,\\
    & y_t - \theta_t^\top x_t \sim\mathcal{N}(0, \exp(a_t)) \,.
\end{align*}
Instead of estimating the state $\theta_t$ with variances fixed {\it a priori}, we estimate both the state and the variances represented by $a_t,b_t$. Although we have removed $\sigma_t^2,Q_t$ as  hyper-parameters, we now have to set priors on $a_0,b_0$ along with the parameters $\rho_a,\rho_b$ controlling the smoothness of the dynamics on the variances.

We apply a bayesian approach. At each step, we start from a prior $p(\theta_{t-1},a_{t-1},b_{t-1}\mid \mathcal{F}_{t-1})$ obtained at the last iteration, where we introduce the filtration of the past observations $\mathcal{F}_t=\sigma(x_1,y_1,...,x_{t-1},y_{t-1})$. Then we obtain a prediction step thanks to the dynamical equations yielding $p(\theta_t,a_t,b_t\mid \mathcal{F}_{t-1})$. Finally Bayes' rule yields the posterior distribution $p(\theta_t,a_t,b_t\mid \mathcal{F}_t)$.

However the posterior distribution is analytically intractable, therefore the principle of VIKING is to apply the classical Variational Bayesian approach \cite{vsmidl2006variational}. The posterior distribution is recursively approximated with a factorized distribution. In our setting we look for the best product $\mathcal{N}(\hat\theta_{t\mid t},P_{t\mid t}) \mathcal{N}(\hat a_{t\mid t},s_{t\mid t})\mathcal{N}(\hat b_{t\mid t},\Sigma_{t\mid t})$ approximating $p(\theta_t,a_t,b_t\mid \mathcal{F}_t)$. The criterion minimized is the Kullback-Leibler (KL) divergence
\begin{align*}
    KL(\mathcal{N}(\hat\theta_{t\mid t},P_{t\mid t}) \mathcal{N}(\hat a_{t\mid t},s_{t\mid t})\mathcal{N}(\hat b_{t\mid t},\Sigma_{t\mid t})\ ||\ p(\theta_t,a_t,b_t\mid \mathcal{F}_t)) \,,
\end{align*}
where $KL(p,q)=\int p\log(p/q)dp$. At each step it yields a coupled optimization problem in the three gaussian distributions. The classical iterative method (see for instance \cite{tzikas2008variational}) consists in computing alternately $\exp(\mathbb{E}[\log p(\theta_t,a_t,b_t\mid \mathcal{F}_t)])$ where the expected value is taken with respect to two of the three latent variables, and identifying the desired first two moments with respect to the other latent variable. However the expression $\exp(\mathbb{E}_{\theta_t,b_t}[\log p(\theta_t,a_t,b_t\mid \mathcal{F}_t)])$ doesn't match a gaussian distribution in $a_t$, and similarly for $b_t$. We therefore use the first two moments of the gaussian distribution to derive an upper-bound of the KL divergence for which we have an analytical solution. We refer to \cite{tsw} for the detailed derivation of the algorithm.

\section{Experiments}\label{sec:experiments}
%We present the results obtained on the competition dataset. 
We display the performance of the introduced methods that we call experts. Then we use aggregation of experts to leverage specificities of each forecaster. The end of the section is devoted to a discussion of our day-to-day strategy during the competition. Finally, we refer to the implementation for more details\footnote{\href{https://gitlab.com/JosephdeVilmarest/state-space-post-covid-forecasting}{https://gitlab.com/JosephdeVilmarest/state-space-post-covid-forecasting}}.

\subsection{Individual Experts}\label{sec:exp_indiv}
We first display in Table \ref{tab:indiv_offline} the results of the statistical and machine learning methods of Section \ref{sec:statsml}, with or without the intraday correction of Section \ref{sec:intraday}. To present the improvement brought by the intraday correction we give the performance during a stable period (after the training of the model, but before the covid crisis). We observe that the only model for which the intraday correction doesn't improve the performance (RF\_GAM) is the one including already a residual correction. The improvement during the evaluation period (2021) is much bigger (57\% decrease of the MAE for the MLP for instance), and it is natural as the intraday correction is an autoregressive, that is, an adaptive model.
\begin{table}
    \begin{center}
    \caption{Mean Average Error (in MW) of each method of Section \ref{sec:data_pres} during normal test period.}
    \label{tab:indiv_offline}
    \begin{tabular}{|c|c|c|c|c|c|c|}
        \hline
        Adaptation & AR & Linear & GAM & RF & RF\_GAM & MLP \\
        \hline
        Offline & 29.3 & 20.8 & 20.7 & 24.6 & 23.0 & 21.2 \\
        Offline intraday & 27.0 & 19.9 & 19.3 & 24.4 & 23.7 & 20.6 \\
        \hline
    \end{tabular}
    \end{center}
    Models are trained up to Jan. 1\textsuperscript{st} 2020 and tested during the next two months before the break of March.
\end{table}

Then we focus on adaptive models to show the improvements due to each setting, see Table \ref{tab:individual}.
\begin{table}[]
    \begin{center}
    \caption{Mean Average Error (in MW) of each method during the competition evaluation set (2021-01-18 to 2021-02-16).}
    \label{tab:individual}
    \begin{tabular}{|c|c|c|c|c|}
        \hline
        Adaptation & AR & Linear & GAM & MLP \\
        \hline
        Offline & 14.6 & 22.8 & 22.7 & 16.7 \\
        \hline
        Static & 20.5 & 15.7 & 17.0 & 22.9 \\
        Static break & 27.9 & 14.4 & 28.4 & 35.4 \\
        Dynamic & 14.4 & 14.9 & 15.3 & 13.0 \\
        Dynamic break & 16.2 & 13.6 & 14.3 & 12.3 \\
        Dynamic big & 14.3 & 11.2 & 12.4 & 12.4 \\
        \hline
        VIKING & 14.4 & 11.5 & 12.7 & 12.5 \\
        \hline
        %Aggregation & 14.9 & 11.4 & 11.6 & 11.9 \\
        %\hline
    \end{tabular}
    \end{center}
    The performances are displayed for each model after intraday correction. As a comparison, re-training the random forest every day yields an online RF of MAE 15.0 MW, and an online GAM\_RF of MAE 18.1 MW.
\end{table}
We have 4 different models (autoregressive, linear, GAM and MLP). For each one, we try the various adaptation settings (no adaptation, Kalman filters and VIKING). Kalman filters with constant covariance matrix proportional to the identity obtain the best results. That is not the case on the data previous to the competition and it depends on the intrinsic evolution of the data.

\begin{figure*}
    \centering
    \includegraphics[width=5cm]{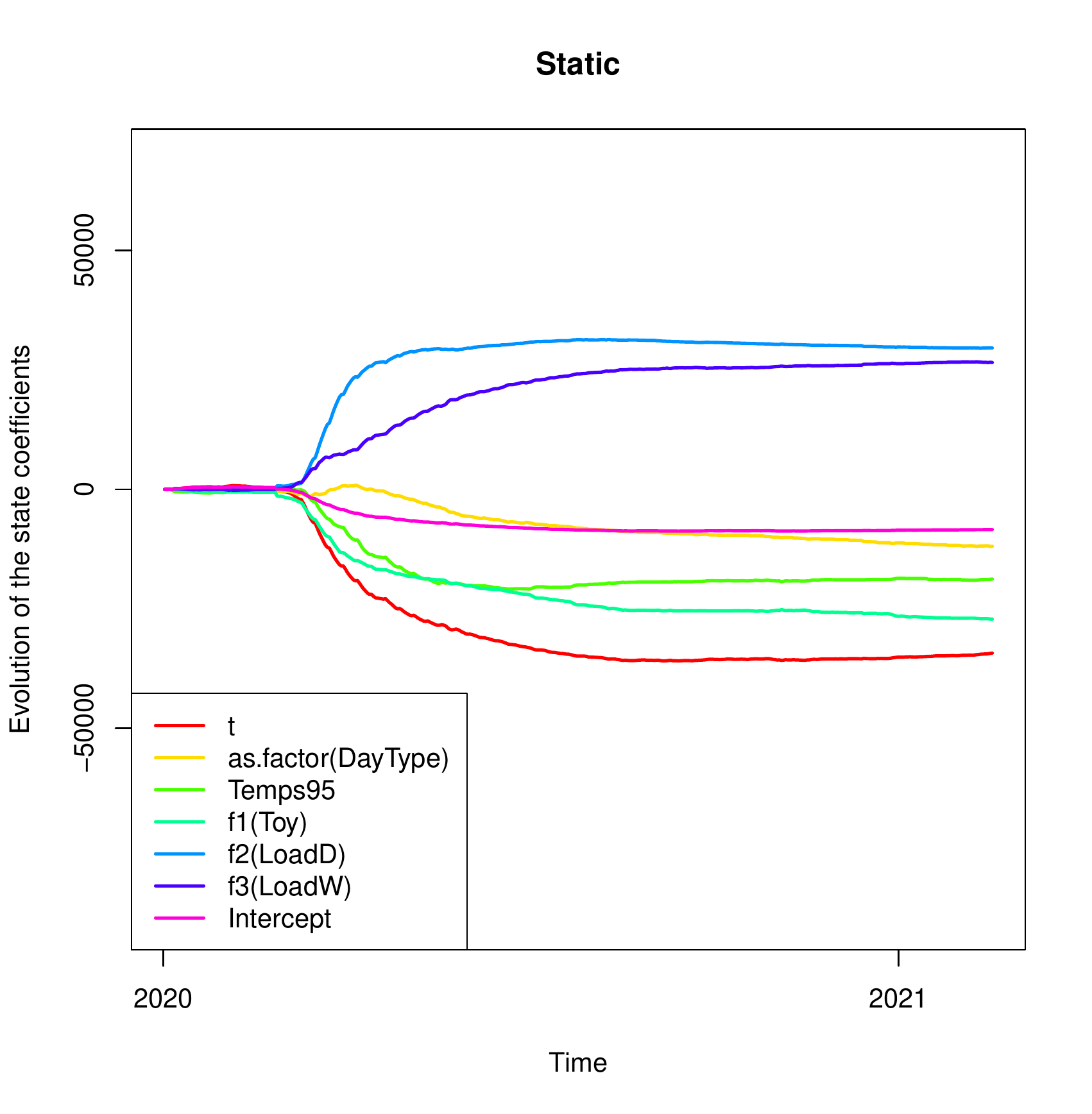}
    \includegraphics[width=5cm]{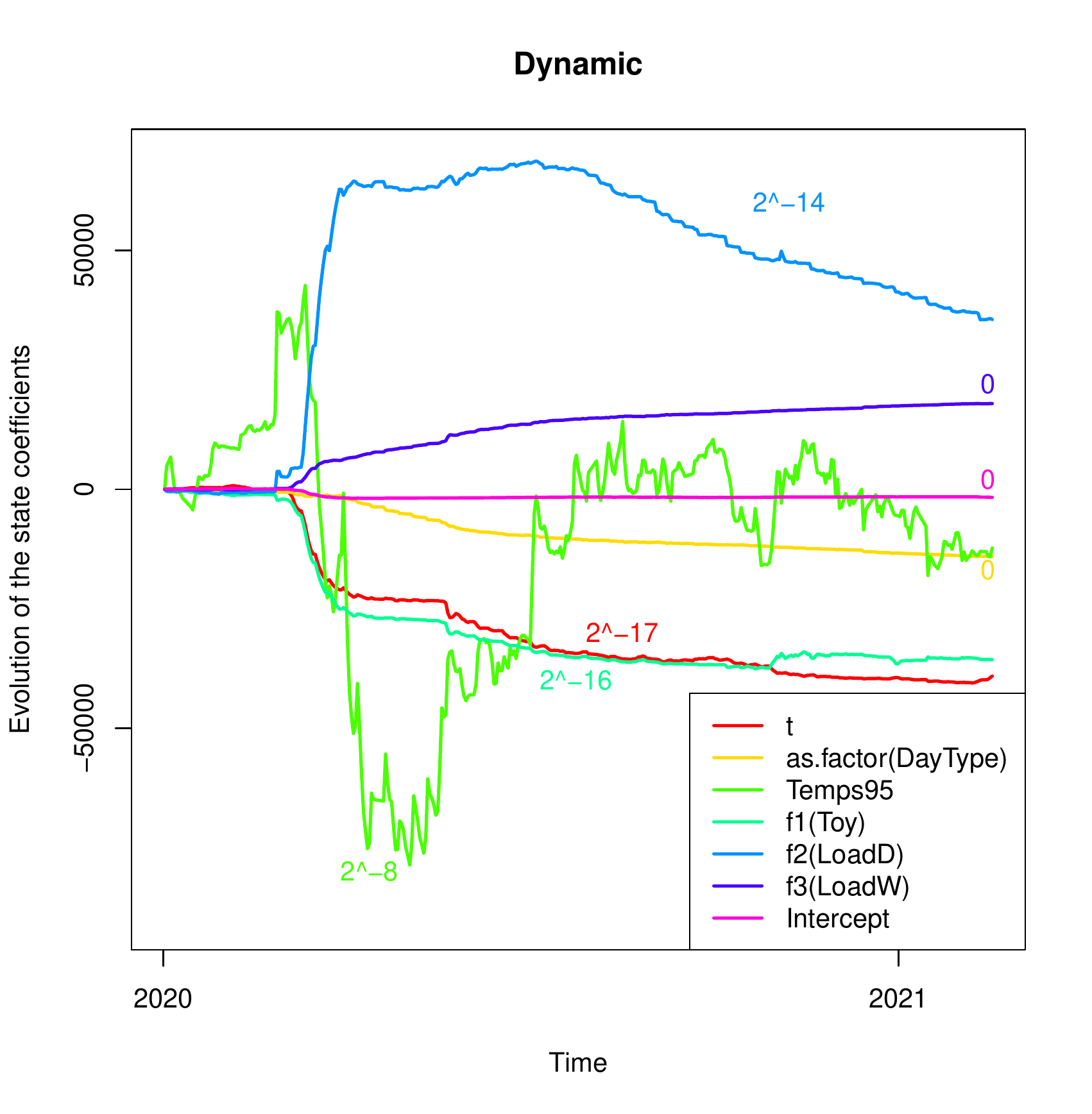}
    \includegraphics[width=5cm]{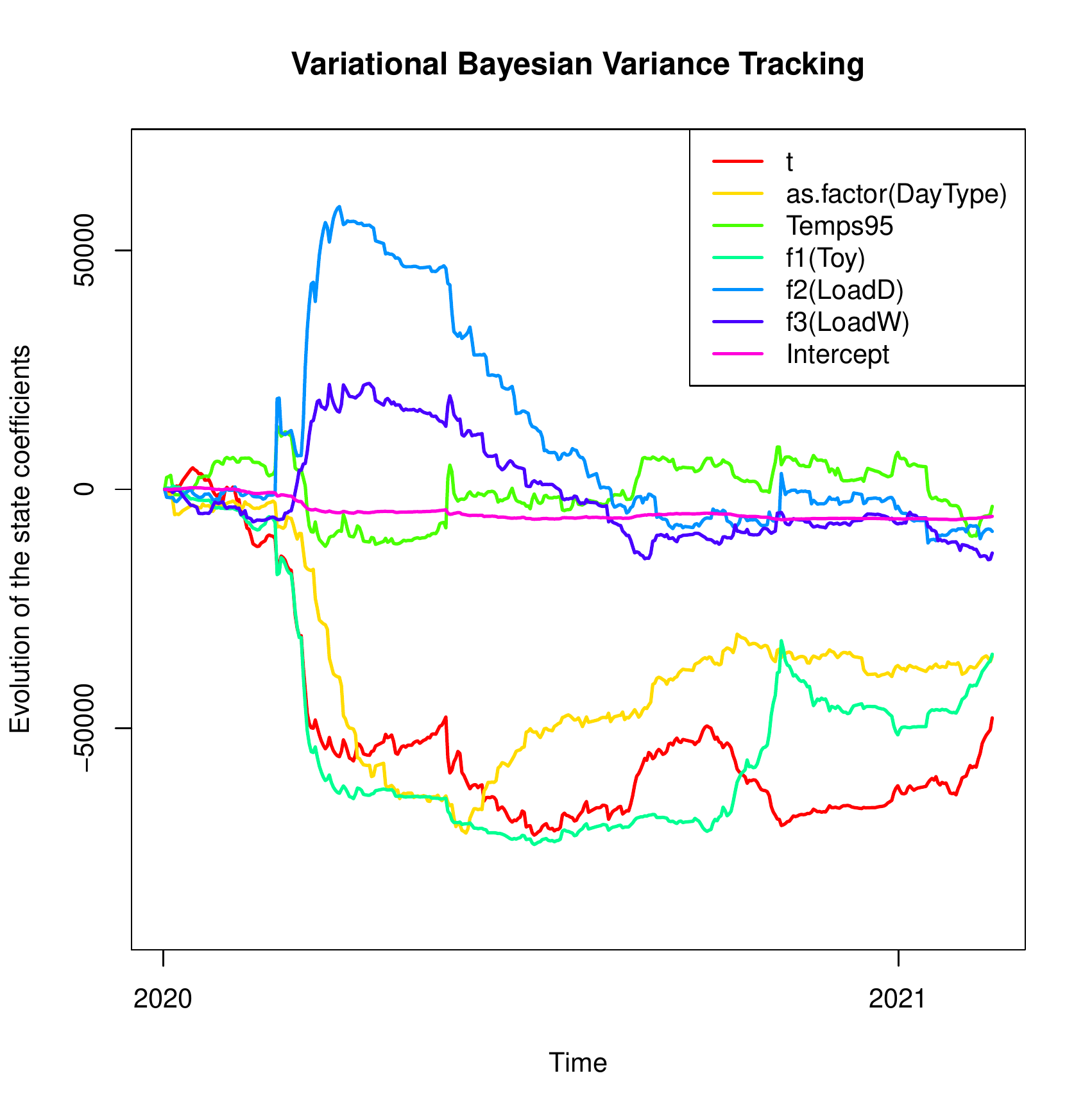}
    \caption{Evolution of the state coefficients for various adaptations of the GAM, see Section \ref{sec:adaptation}. On the left, we use the Kalman Filter in the static setting (degenerate covariance matrix $Q_t=0$). On the middle, the dynamic setting where the variances are constant, and we provide the ratio $Q/\sigma^2=diag(2^{-17},0,2^{-8},2^{-16},2^{-14},0,0)$: we observe that the coefficient corresponding to the biggest coefficient of $Q$ (the effect of $Temps95$) evolves much faster. On the right, the VIKING setting where we estimate the variances adaptively.}
    \label{fig:evol_gam}
\end{figure*}

We illustrate the different settings in Figure \ref{fig:evol_gam} where we display the evolution of the state coefficient for the GAM adaptation strategies.

\subsection{Aggregation}
Online robust aggregation of experts \cite{Cesa-Bianchi:2006} is a powerful model agnostic approach for time series forecasting, already applied to load forecasting during the lockdown in \cite{obst2021adaptive}.
We use the ML-Poly algorithm proposed in \cite{gaillard2014second} and implemented in the \texttt{R} package \texttt{opera} \cite{gaillard2016opera} to compute these online weights.

\begin{figure}
    \centering
    \includegraphics[height=6cm]{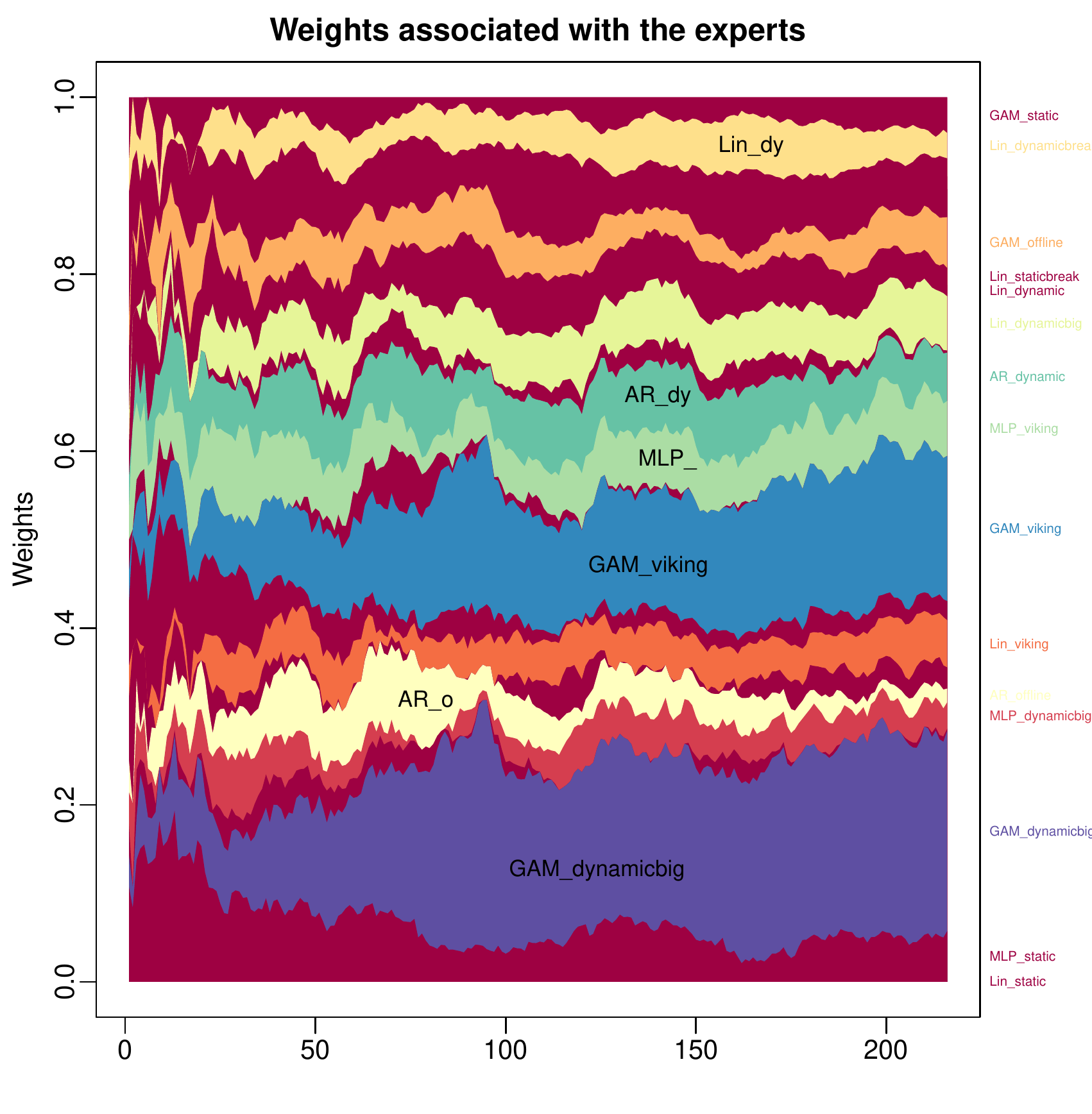}
    \caption{Evolution of the aggregation weights at 3PM from July 1\textsuperscript{st} 2020 to February 16\textsuperscript{th} 2021.}
    \label{fig:weights}
\end{figure}

The aggregation weights are estimated independently for each hour of the day. We summarize different variants in Table \ref{tab:aggregation}.
\begin{table}[]
    \centering
    \caption{Mean average error of aggregation strategies (in MW) during the competition evaluation set (2021-01-18 to 2021-02-16).}
    \label{tab:aggregation}
    \begin{tabular}{|c|c|c|c|c|c|}
        \hline
        Adaptation & AR & Linear & GAM & MLP & All \\
        \hline
        Best expert & 14.3 & 11.2 & 12.7 & 12.3 & 11.2 \\
        \hline
        Aggregation & 14.4 & 11.4 & 11.7 & 11.9 & {\bf 10.9} \\
        \hline
    \end{tabular}
\end{table}
First, for each family of models we compute the aggregation of all the adaptation settings (7 for each). Then we aggregate all of them (28 models). An example of the weights obtained at 3PM is displayed in Figure \ref{fig:weights}.
The aggregation presented in this paper obtains a performance close to our strategy winning the competition (degradation of about 0.05 MW).

\subsection{Day-to-day Forecasts}\label{sec:daybyday}
During the competition our predictions were not exactly the ones of the aggregation method presented in the previous subsection. There are mainly two reasons for that. 

\begin{figure}
    \centering
    \includegraphics[width=8cm]{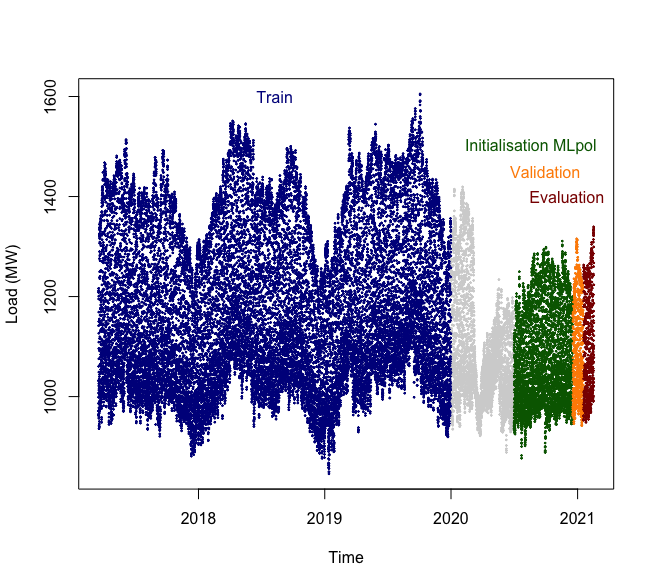}
    \caption{Segmentation of the data set. Meteorological corrections as well as time-invariant forecasts were trained on the train period (up to January 1\textsuperscript{st} 2020). Adaptive forecasting methods were evolving on the whole period. Then the aggregation weights were trained by MLpol from July 1\textsuperscript{st} 2020, and the expert selection was determined with respect to the validation set.}
    \label{fig:segmentation}
\end{figure}

\begin{figure}
    \centering
    \includegraphics[height=6cm]{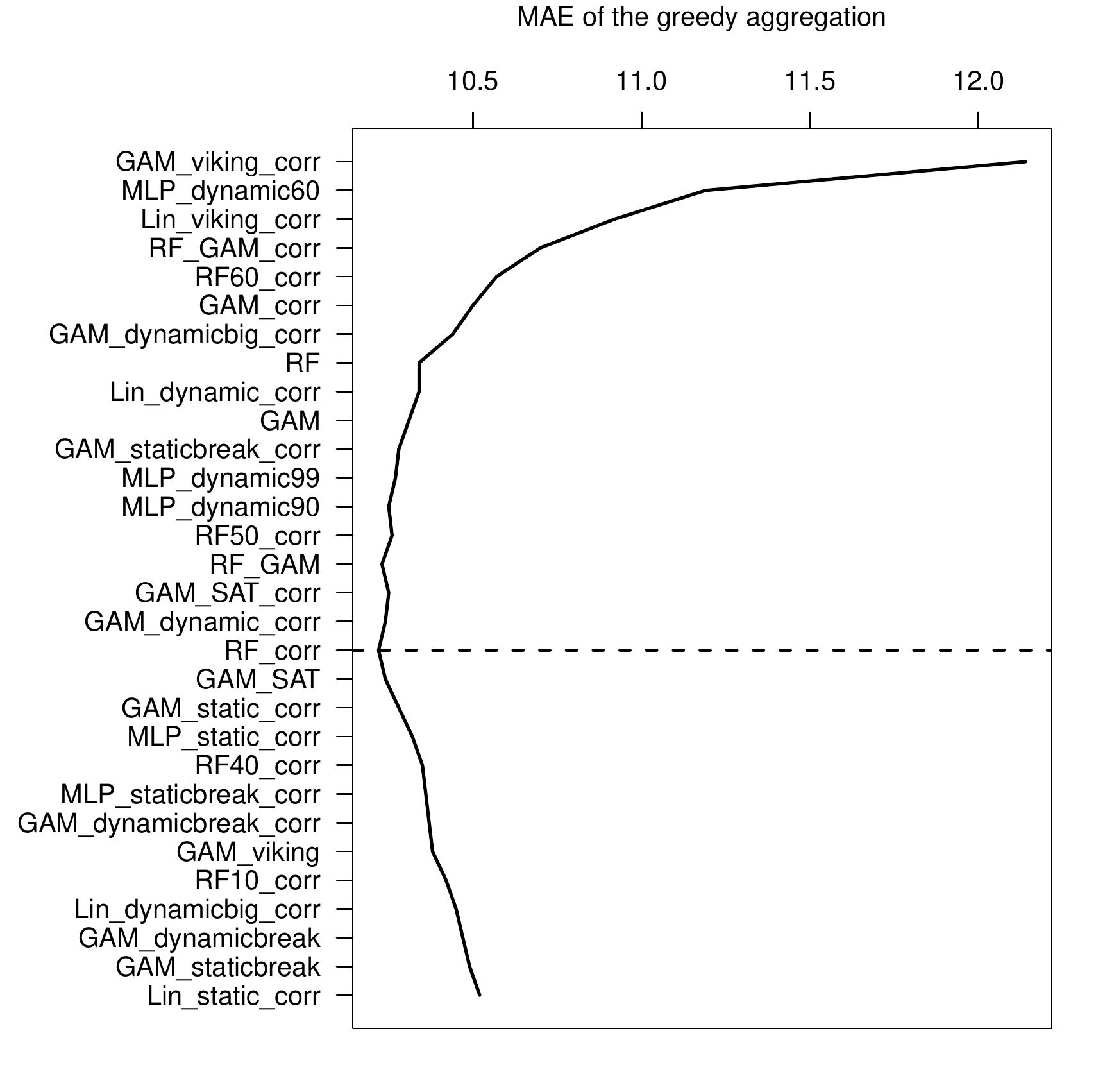}
    \caption{Evolution of the validation MAE as the expert selection grows from 1 to 30 experts. The nomenclature is provided in the Appendix.}
    \label{fig:greedy_selection}
\end{figure}

First, we considered a bigger set of forecasting methods (we had $72$ experts). It seemed reasonable to prune the strategy for the sake of clarity of paper at the cost of a very small change of error, but it is interesting to present also the predictions used during the evaluation. We found a trade-off in the selection of experts. Indeed too much experts in the aggregation yields poor performances. We applied a greedy procedure to select the experts we keep in the aggregation: we begin with an empty set, and at each step we add the one improving the most the performance. That performance was evaluated with the MAE on the last month of the training data set. We provide in Figure \ref{fig:segmentation} a graphical representation of how we defined different time periods. We refer to Figure \ref{fig:greedy_selection} for the evolution of the validation MAE as the selection grows. We observe a sharp decrease of the error as experts are added with high diversity and then a slow increase of the loss as the set of experts becomes too large.

Second, we were constantly experimenting the different strategies. We used a variational bayesian method that was a prior version of the one of this paper. We also changed a lot the aggregation procedure.

We refer to the appendix for a detailed presentation of our daily strategy. Overall these day-by-day changes degraded the performance, if we had stayed on the first strategy with no change at all, our MAE would have been 10.51 MW instead of 10.84 MW. The critical issue in such unstable periods is to find the right validation period to select the prediction procedure. The month before the evaluation period seems {\it a posteriori} a good compromise. During the competition we changed "manually" based on the performances in a shorter range, for instance considering an expert performing well on the last few weeks for a specific day type ... We should have trusted the aggregation's robustness.

\section{Conclusion}
In this paper we presented our procedure to win a competition on electricity load forecasting during an unstable period. Our approach relies heavily on state-space models and the competition was the first data set on which was applied a recent approach to adapt the variances of a state-space model. Some perspectives have been raised during the competition such as interpretability of the global approach and a better understanding of the error propagation along the different adaptations (intraday correction, Kalman filtering, variance tracking and aggregation).

Finally, similar state-space methods have been applied to obtain the first place in another competition in which the objective was to forecast the electricity consumption of a building\footnote{\href{http://www.gecad.isep.ipp.pt/smartgridcompetitions/}{http://www.gecad.isep.ipp.pt/smartgridcompetitions/}}.

\section*{Appendix}
\subsection{Nomenclature}
The experts \texttt{AR}, \texttt{Lin}, \texttt{GAM}, \texttt{RF}, \texttt{RF\_GAM}, \texttt{MLP} are the ones presented in that same order in Section \ref{sec:statsml}.

Names of the form \texttt{model\_setting} refer to the expert obtained by state-space adaptation of the model \texttt{model} with the setting \texttt{setting}. For instance, \texttt{Lin\_dynamic} refers to a linear model adapted with the Kalman filter in the dynamic setting, {\it c.f.} Section \ref{sec:kf}.

We consider quantile variants of \texttt{RF}, denoted by \texttt{RFq} where \texttt{q} is the quantile order in percent ({\it e.g.} \texttt{RF40} is the quantile random forest of quantile value 0.4). We also consider a quantile variant of the dynamic MLP denoted by similar names (\texttt{MLP\_dynamic60} is the quantile 0.6 of the MLP in the dynamic setting).

Furthermore, we introduce an expert named \texttt{GAM\_SAT} forecasting each day with the GAM as if it were a Saturday motivated by \cite{obst2021adaptive}.

Finally, each expert \texttt{x} yields another expert \texttt{x\_corr} after intraday correction.

\subsection{Day-to-day Evolution of the Forecasting Strategy}
As explained in Section \ref{sec:daybyday}, our strategy evolved in time and we recall here every change.
\begin{itemize}
\item
\textbf{From January 18\textsuperscript{th} to January 24\textsuperscript{th}}: we used the following set of experts obtained by the greedy selection described in Section \ref{sec:daybyday}: \texttt{RF}, \texttt{RF\_corr}, \texttt{RF50\_corr}, \texttt{RF60\_corr},  \texttt{Lin\_dynamic\_corr}, \texttt{Lin\_viking\_corr}, \texttt{GAM}, \texttt{GAM\_corr}, \texttt{GAM\_staticbreak\_corr}, \texttt{GAM\_dynamic\_corr}, \texttt{GAM\_viking\_corr}, \texttt{GAM\_dynamicbig\_corr}, \texttt{RF\_GAM}, \texttt{RF\_GAM\_corr}, \texttt{GAM\_SAT\_corr}, \texttt{MLP\_dynamic60}, \texttt{MLP\_dynamic90}, \texttt{MLP\_dynamic99}. We aggregated with ML-poly with an aggregation estimated independently for each hour, with the absolute loss. We found afterwards a bug in \texttt{RF50\_corr, RF60\_corr}: the quantile RF were set to 0 on the test set so that these two experts were simple intraday autoregressive trained in an unwanted manner.

\item
\textbf{From January 25\textsuperscript{th} to January 31\textsuperscript{st}}: we removed the experts \texttt{RF50\_corr, RF60\_corr} and we replaced them with \texttt{AR\_corr}.

\item
\textbf{February 1\textsuperscript{st} and 2\textsuperscript{nd}}:
we used the uniform average between three forecasts. First, the previous aggregation. Second, another aggregation procedure called RF-stacking, consisting in a quantile random forest minimizing the MAE and taking as input the 72 experts as well as the day type and hour of the day. Third, a benchmark close to the one given by the competition organizers: we predict each time with the last available load of the same hour and the same day group (week days, saturdays and sundays).

\item
\textbf{From February 3\textsuperscript{rd} to 7\textsuperscript{th}}: we removed the benchmark which damaged the performances. For \textbf{Feb. 5\textsuperscript{th}} we corrected the ML-poly prediction using a special day correction, once we observe that Feb. 5\textsuperscript{th} had a special behaviour in the last three years. Precisely, we observed that the relative error of the model is significantly negative on the last three years, a behaviour that may come from a bank holiday for instance. Therefore we fit a smoothed function of the time of day on the relative error and we applied it to our forecast. We truncated so that there is no correction during night. See the shape of the correction in Figure \ref{fig:specialday}.
\begin{figure}
    \centering
    \includegraphics[width=7cm]{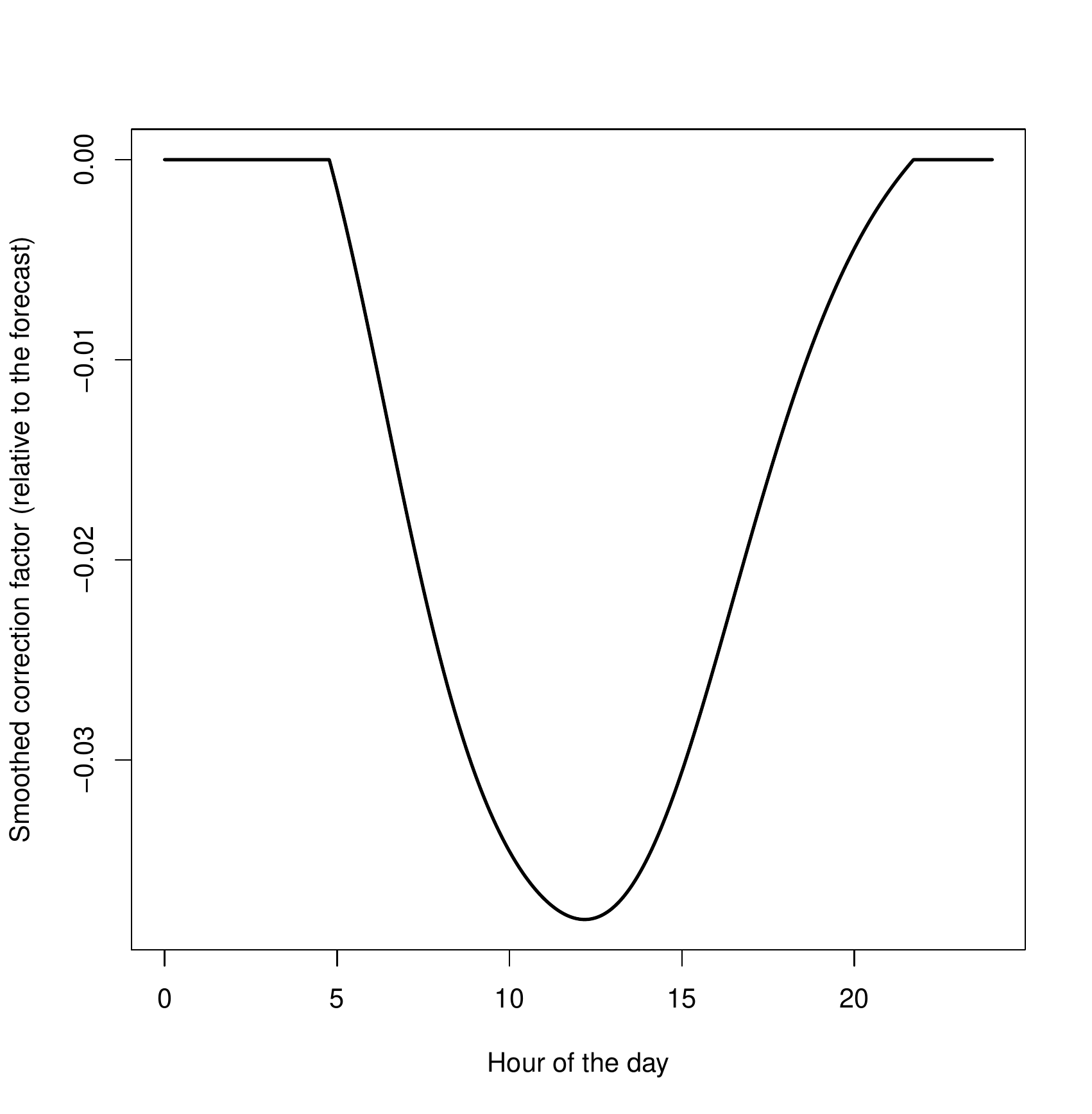}
    \caption{Special day correction applied on February 5\textsuperscript{th}. It is a multiplicative correction, {\it e.g.} at midday we reduce our forecast by about 3.8\%.}
    \label{fig:specialday}
\end{figure}

\item
\textbf{February 8\textsuperscript{th}}: we used the single expert \texttt{Lin\_dynamicbig\_corr} as we observed that it was by far our best expert on the last week, and it seemed to perform especially well on Mondays.

\item
\textbf{February 9\textsuperscript{th}}: we came back to the average between ML-poly and the RF-stacking but we added to the aggregation ML-poly the expert \texttt{Lin\_dynamicbig\_corr}, and we replaced the expert \texttt{AR\_corr} with another expert \texttt{AR\_intra} incorporating directly the intraday correction in the autoregressive, instead of correcting an autoregressive based only on daily lags.

\item
\textbf{February 10\textsuperscript{th} and 11\textsuperscript{th}}: we removed the RF-stacking which degraded our performances since its introduction and we kept only the aggregation ML-poly.

\item
\textbf{February 12\textsuperscript{th} and 13\textsuperscript{th}}:
we corrected {\it a posteriori} the electricity load for February 5\textsuperscript{th} with the special day correction. It was important to do it on that day as the weekly lags is important in the models.

\item
\textbf{February 14\textsuperscript{th}}: we used once again the average between the ML-poly aggregation and the RF-stacking, as we observed that the RF-stacking is especially good on Sunday.

\item
\textbf{February 15\textsuperscript{th} and 16\textsuperscript{th}}: we used only the ML-poly aggregation.
\end{itemize}

%\section*{Acknowledgment}

\bibliographystyle{IEEEtran}
\bibliography{IEEEabrv,biblio.bib}

\end{document}